\documentclass[11pt,a4paper]{article}

\usepackage[T1]{fontenc}
\usepackage[utf8]{inputenc}
\usepackage{lmodern}
\usepackage{microtype}
\usepackage[margin=2.0cm]{geometry}
\usepackage{amsmath,amssymb}
\usepackage{booktabs,tabularx}
\usepackage{graphicx}
\usepackage{listings}
\usepackage{tikz}
\usetikzlibrary{arrows.meta, positioning, shapes.geometric, backgrounds}
\usepackage[numbers,sort&compress]{natbib}
\usepackage{url}
\usepackage[hidelinks]{hyperref}
\title{Beyond Surface Metrics: Multi-Condition Requirements in Enterprise Maintenance and AI Coding Benchmarks}
\author{Heeyong Eun\\KPMG Korea, Seoul, South Korea\\\texttt{heeyonge@gmail.com}}
\date{}

\begin{document}
\maketitle

\begin{abstract}
Repository-level AI coding benchmarks measure whether agents solve sampled software tasks, but task size and benchmark success do not directly describe the conditions that constitute maintenance completion. We examine \textbf{multi-condition fulfillment}: situations in which a maintenance activity must satisfy several operational conditions, sometimes linked by directional dependencies. Using a frozen 18-criteria taxonomy, we analyze 2,925 Apache Fineract issues and 944 AI benchmark tasks (SWE-bench Pro, SWE-PolyBench Java, and SWE-EVO) through three complementary lenses: attribute breadth ($|B|$), corpus-level co-occurrence ($\phi$), and source-grounded directional requirement relationships. First, 1,177 benchmark task pairs matched on issue classification, modified-file count, and code churn have mean fulfillment-set Jaccard distance $0.8138$; 57.4\% have disjoint condition sets. Second, attribute count does not explain the structural contrast: SWE-bench Pro averages $|B|=2.47$ versus Fineract $1.78$. Temporal ($B_4$) and numerical ($B_2$) conditions have $\phi=+0.430$ in Fineract versus $-0.013$ in the pooled AI corpus; among $B_4$-tagged tasks, $B_2$ is also specified in 48.85\% (254/520) versus 5.81\% (5/86), respectively. These associations are descriptive, not causal. Third, 16 source-grounded Fineract requirement relationships illustrate directional obligations involving time, calculations, and state preservation. Four selected benchmark cases then illustrate how parts of such fulfillment contracts appear in inspected evaluation assertions; they do not estimate benchmark-wide oracle omissions. Our results show that matched surface task profiles can conceal disjoint fulfillment-condition sets, while condition counts alone do not capture observed co-specification patterns. We therefore advocate reporting \textbf{condition breadth, co-occurrence, and documented directional relationships} as distinct properties of maintenance benchmarks. The corpus contrast is descriptive and may reflect domain differences; no production failure rate or causal selection mechanism is inferred.
\end{abstract}

\noindent\textbf{Keywords:} AI Coding Agents; Software Maintenance; Benchmark Evaluation; Multi-Condition Fulfillment; Requirement Dependencies; Enterprise Software

\section{Introduction}

\subsection{Motivation: What Counts as Maintenance Completion?}
The rapid evolution of Large Language Models (LLMs) and autonomous scaffolding frameworks has transformed automated program repair into repository-scale software engineering. Benchmarks such as SWE-bench \cite{jimenez2024swebench}, SWE-bench Pro \cite{deng2025swebenchpro}, SWE-PolyBench \cite{rashid2025swepolybench}, and SWE-EVO \cite{thai2025sweevo} evaluate coding agents on tasks derived from software issues, repository changes, or evolution histories. Although coding-agent benchmarks provide quantitative performance measurements, their leaderboard scores do not directly establish readiness for enterprise software maintenance.

A fundamental question precedes test adequacy: \textit{what combination of conditions defines completion of a maintenance task?} In enterprise services, a single change can entail temporally ordered recalculation, preservation of prior data, and consistent downstream state. These are not necessarily independent checklist items: one operation can trigger another, and the task is complete only when the applicable conditions hold together. We call this \textbf{multi-condition fulfillment}, distinguishing the \emph{number} of conditions from their \emph{relationships}. Apache Fineract backdating requirements provide concrete instances of such linked obligations; whether comparable structures are visible in benchmark task descriptions is an empirical question.

The distinction matters for benchmark interpretation. A task may mention several conditions without stating how they interact, while a compact issue may specify a sequence of dependent operations. Evaluation tests are one downstream way to check those conditions, not the definition of the conditions themselves. We therefore first compare requirement structures across corpora, and only then inspect selected benchmark tests as illustrative consequences for evaluation design.

\subsection{Limitations of Surface Complexity Metrics}
Existing benchmark frameworks predominantly evaluate task difficulty through surface syntactic and navigational metrics: (1)~\textbf{Syntactic Churn} (LOC added/deleted, file counts); (2)~\textbf{Repository Topography} (search space breadth, AST exploration radius, directory depth \cite{shayanfar2025snc}); and (3)~\textbf{Generic Labels} ("Bug Fix", "Feature").

We examine whether surface task descriptors obscure differences in \textit{fulfillment-condition structures}. Some maintenance activities ($A$) impose dependent requirements ($B$), as illustrated by backdated financial operations:
\begin{equation}
A \xrightarrow{\Delta S} B_{\text{temporal}} \xrightarrow{\text{triggers}} B_{\text{numeric}} \models B_{\text{invariant}}
\end{equation}
The analytical question is whether tasks with similar surface profiles describe comparable sets and relationships of fulfillment conditions; evaluation coverage is a subsequent implication, not the primary outcome measured here.

\subsection{Research Questions}
This investigation is guided by two fundamental research questions:
\begin{itemize}
  \item \textbf{RQ1 (Semantic Heterogeneity \& Attribute Diversity)}: To what extent do conventional benchmark task categories and surface churn metrics conceal semantic heterogeneity, and how does the diversity of required maintenance attributes ($|B|$) compare between enterprise maintenance and AI coding benchmarks?
  \item \textbf{RQ2 (Multi-Condition Fulfillment Structure)}: (a)~\textit{Corpus-Level Co-occurrence (RQ2a)}: How do observed patterns of jointly specified conditions ($\phi$) differ between the studied enterprise and benchmark corpora? (b)~\textit{Directional Dependencies (RQ2b)}: What source-grounded action--condition and condition--condition relationships characterize selected enterprise maintenance requirements, and what do selected benchmark cases illustrate about representing these obligations in evaluation?
\end{itemize}

\subsection{Core Contributions}
We present a mixed-methods empirical study grounded in verified raw data across three distinct analytical levels:
\begin{enumerate}
  \item \textbf{Unmasking Semantic Heterogeneity (Level 1 / E1)}: In 1,177 controlled task pairs matched on repository native types and code churn (714 SWE-bench Pro pairs, 463 PolyBench Java pairs, 581 unique tasks), matching on surface task descriptors does not guarantee similar fulfillment-condition sets (mean $d_J = 0.8138$, median $1.0000$, $\sigma = 0.2745$), with 57.4\% of pairs (676 pairs) exhibiting completely disjoint criteria ($d_J = 1.0$).
  \item \textbf{Attribute Diversity Paradox (Level 2 / E2 Macro)}: Across 944 AI tasks (SWE-bench Pro $N=731$, PolyBench $N=165$, SWE-EVO $N=48$), SWE-bench Pro exhibits higher attribute diversity (mean $|B|=2.47$) than enterprise core banking (Fineract $1.78$, Welch's $t=15.78, p < 10^{-50}$). However, linear co-occurrence between temporal backdating ($B_4$) and numerical recalculation ($B_2$) in AI benchmarks shows near-zero correlation ($\phi = -0.013, p=0.825$, 5 tasks) vs. enterprise banking ($\phi = +0.430, p < 10^{-90}$, 254 issues), showing that attribute breadth alone does not characterize joint-condition structure.
  \item \textbf{Directional Fulfillment Contracts (Level 3 / E2 Micro)}: We trace 16 source-grounded Fineract requirement relationships (19 quotes) to illustrate how operations such as backdating, recalculation, and state preservation form joint completion obligations. Four selected benchmark cases illustrate possible alignment boundaries between explicit conditions and inspected assertions; they do not establish a population-level failure rate.
  \item \textbf{Reporting Implications}: We distinguish attribute breadth, statistical co-occurrence, and source-grounded directionality as non-interchangeable properties of maintenance tasks, with secondary implications for specification-aligned benchmark checks.
\end{enumerate}

\section{Related Work \& Research Positioning}

\subsection{Evolution of AI Coding Benchmarks}
Automated code generation benchmarks have evolved rapidly from isolated synthetic function completion (e.g., HumanEval \cite{chen2021codex}, MBPP) to repository-level software maintenance. SWE-bench \cite{jimenez2024swebench} pioneered the extraction of historical pull requests from Python repositories, packaging them with reproducible execution containers. Subsequent benchmarks extended this paradigm: (1)~\textbf{SWE-PolyBench} \cite{rashid2025swepolybench} supports repository-level evaluation across Java, JavaScript, TypeScript, and Python; (2)~\textbf{SWE-bench Pro} \cite{deng2025swebenchpro} targets long-horizon, multi-file software engineering tasks across public, held-out, and commercial splits; and (3)~\textbf{SWE-EVO} \cite{thai2025sweevo} introduces longitudinal software evolution across sequential commits to evaluate regression avoidance. Across all these benchmarks, candidate patches are evaluated by executing automated test harnesses to verify whether target unit tests transition from failing to passing (\texttt{fail\_to\_pass}) without inducing regressions in pre-existing tests (\texttt{pass\_to\_pass}).

\subsection{Characterizing Benchmark Complexity \& Audit Studies}
Several empirical frameworks have investigated the demands of repository-level coding tasks. \textbf{Syntactic \& Navigational Complexity (SNC)} \cite{shayanfar2025snc} profiles Spread (file distribution entropy), Novelty (the proportion of added versus deleted code), and Centrality (dependency graph importance), examining task demands through code-change and repository structure. In contrast, \textbf{UTBoost} \cite{yu2025utboost} identifies insufficient issue-specific benchmark tests that can allow incorrect candidate patches to pass, and augments tests to improve patch discrimination. These studies examine repository-level demands and patch evaluation, respectively, rather than the specified structure of multi-condition maintenance completion.

\textbf{The Distinct Dimension: Multi-Condition Fulfillment Structure.} Test-adequacy studies such as UTBoost address whether existing tests discriminate correct and incorrect patches. We instead examine what a maintenance task \emph{requires}: which completion conditions are specified, which appear together, and which are directionally dependent. We do not claim that test coverage has been unexplored, nor that our descriptive corpus comparisons prove a missing condition in any entire benchmark harness.

\subsection{Research Positioning: From Task Size to Joint Fulfillment Conditions}
As summarized in Table~\ref{tab:positioning}, repository-complexity studies analyze properties of code changes and navigation demands, while test-audit work examines patch evaluation. Our unit of analysis is the \textbf{maintenance fulfillment contract}: developer activities ($A$), specified conditions ($B$), and, where supported by source text, relationships among those conditions. We distinguish three complementary observations: (1)~condition breadth within an issue ($|B|$); (2)~corpus-level co-occurrence ($\phi$), which does not establish direction; and (3)~explicit or source-grounded directional obligations linking actions to downstream states. Selected oracle inspections are a \emph{secondary application} of this characterization.

The principal research question is therefore not simply whether a test catches a buggy patch. It is: \textit{Which fulfillment conditions co-occur across the analyzed corpora, and what source-grounded dependencies characterize completion in selected enterprise maintenance requirements?}
\begin{table}[htbp]
\centering
\footnotesize
\renewcommand{\arraystretch}{0.92}
\caption{Research positioning by unit of analysis and primary contribution.}
\label{tab:positioning}
\begin{tabularx}{\linewidth}{p{2.5cm} X X X}
\toprule
\textbf{Analytical Dimension} & \textbf{SNC Audit \cite{shayanfar2025snc}} & \textbf{UTBoost \cite{yu2025utboost}} & \textbf{This Work} \\
\midrule
\textbf{Primary Problem} & 
Repository topography ambiguity and task demand categorization & 
Incomplete tests and false pass judgments & 
Multi-condition structure of maintenance completion \\
\addlinespace
\textbf{Unit of Analysis} & 
Repository graph metrics (AST distance, spread, novelty) & 
Tests and candidate patches in SWE-bench tasks & 
Requirement fulfillment invariants ($A \rightarrow \Delta S \rightarrow B$) \\
\addlinespace
\textbf{Requirement Perspective} & 
Surface issue labels (Bug, Feature, Enhancement) & 
Task-specific tests and patch evaluation & 
Ontological taxonomy (ISO 14764 / 25010) with verified quotes \\
\addlinespace
\textbf{Oracle Evaluation Scope} & 
Accepts existing benchmark test harnesses as ground truth & 
Augments benchmark tests to flag incorrect passes & 
Secondary case-level assertion inspection \\
\addlinespace
\textbf{Empirical Scope} & 
SWE-bench Lite/Verified (Python repository graphs) & 
SWE-bench Python repos (LLM-augmented test suites) & 
Analyzed AI tasks ($N=944$) vs. Fineract issues ($N=2,925$) \\
\bottomrule
\end{tabularx}
\end{table}

\section{Empirical Foundations \& Methodology}

\subsection{Reference Corpora}
Our empirical foundation comprises two extensive software maintenance corpora:
\begin{itemize}
  \item \textbf{Enterprise Reference Corpus (Apache Fineract)}: Apache Fineract is an open-source core banking platform maintained by the Apache Software Foundation \cite{apachefineract}, providing financial service infrastructure (loan scheduling, multi-currency ledgers, portfolio management). We analyze a collected corpus of \textbf{2,925 maintenance issues} from its issue tracker, comprising 3,811 granular requirement units, including 2,393 issues identified as resolved in the source metadata. Fineract serves as a domain-specific enterprise reference corpus in which maintenance issues can impose financial calculations, audit requirements, and state-consistency constraints; its banking domain limits direct generalization to other systems.
  \item \textbf{AI Coding Benchmark Corpus (944 Analyzed Tasks)}: We analyze \textbf{944 tasks} from three benchmark sources: (1)~\textbf{SWE-bench Pro public V1}: 731 tasks (77.44\%) from its original public split; (2)~\textbf{SWE-PolyBench Java}: 165 tasks (17.48\%) from its Java split; and (3)~\textbf{SWE-EVO}: 48 tasks (5.08\%) included in the analyzed corpus. These counts define the scope of our study rather than the complete populations of all benchmark releases. Across these 944 tasks, \textbf{3,446 itemized criteria assignments} were systematically annotated with exact text quotes and code diff references.
\end{itemize}

\subsection{The Immutable 18-Criteria Maintenance Taxonomy: Provenance and Grounding}
To support a transparent and consistently applied cross-domain comparison, we establish an immutable 18-criteria maintenance taxonomy in three operational stages: (1)~\textbf{Standard-Based Concept Derivation}: Development Activities ($A_1$–$A_6$) were adapted from the maintenance categories of \textbf{ISO/IEC/IEEE 14764} \cite{iso14764} (Corrective, Adaptive, Perfective, Preventive), while Fulfillment Conditions ($B_1$–$B_{12}$) were adapted from the quality characteristics of \textbf{ISO/IEC 25010 (SQuaRE)} \cite{iso25010} formalized as state invariants; (2)~\textbf{Empirical Concretization \& Codebook Freezing}: Through inductive thematic analysis across 2,925 Apache Fineract maintenance issues (3,811 requirement units), standard categories were operationalized into concrete criteria and formally frozen into an immutable codebook (\texttt{framework-snapshot.json}); and (3)~\textbf{Uniform Cross-Corpus Application}: This frozen taxonomy was applied without alteration across the 944 analyzed AI benchmark tasks, maintaining consistent operational definitions across the selected corpora. The proposed 18-criteria taxonomy is an empirically operationalized classification informed by ISO/IEC/IEEE 14764 and ISO/IEC 25010, rather than a direct reproduction of either standard. The six activity categories and twelve fulfillment conditions were developed through inductive analysis of Apache Fineract maintenance requirements, with their operational definitions and classification boundaries documented in the accompanying codebook.

\subsubsection{Ontological Separation: Activities (\texorpdfstring{$A$}{A}) vs. Conditions (\texorpdfstring{$B$}{B})}
Conventional benchmarks collapse maintenance tasks into 1-dimensional categorical labels ("Bug Fix", "Feature"). However, in enterprise service maintenance, the developer's maintenance activity (Activity, $A$) is ontologically distinct from the multi-step system invariants (Conditions, $B$) required to declare the task complete:
$A \xrightarrow{\Delta S} B_{\text{temporal}} \xrightarrow{\text{triggers}} B_{\text{numeric}} \models B_{\text{invariant}}$.
The taxonomy establishes 6 Development Activities ($A_1$--$A_6$: Product Operation, Structure/Refactoring, Testing/QA, Execution/Operations, Deployment/Migration, Process/Docs) and 12 Fulfillment Conditions ($B_1$--$B_{12}$: Functional Result, Numerical Precision, Data Consistency, Temporal Semantics, State Transition, Legacy Preservation, Concurrency Boundaries, Authorization, Performance, Service Continuity, Interface Compatibility, User Interaction), quantified across corpora in Table~\ref{tab:stratified_distribution}.

\subsubsection{Annotation Procedure, Semantic Validity, and Boundary Resolution}
To ensure criteria assignments reflect authentic operational semantics rather than keyword matching, we enforced strict annotation protocols: (1)~\textbf{Operational Rules}: $A$ labels required concrete code diff intent ($A_1$ functional logic, $A_2$ structural refactoring), while $B$ conditions required explicit specification mandates for state invariants or quality constraints; (2)~\textbf{Boundary Adjudication}: (a)~\textit{$B_1$ vs. $B_2$}: input validation and boolean returns were coded as $B_1$, whereas financial calculations, rounding, or multi-threaded counters were coded as $B_2$; (b)~\textit{$A_1$ vs. $A_2$}: functional behavior modifications were classified as $A_1$, whereas internal refactoring preserving behavior was classified as $A_2$; (c)~\textit{$B_3$ vs. $B_6$}: multi-entity referential integrity was coded as $B_3$, whereas legacy data retention and non-regression guarantees were coded as $B_6$; (3)~\textbf{Batch Protocol \& Verification}: Semantic coding across the 944 tasks was conducted by the primary investigator via an iterative batch protocol binding all 3,446 criteria assignments to verbatim textual quotes and rationales. An automated pipeline (\texttt{validate\_*.py}) programmatically verified contiguous task sequences, file hashes, exact string offsets in raw JSON, and codebook schema compliance, ensuring quotation fidelity rather than dual-coder consensus (documented in Section~\ref{sec:threats}); and (4)~\textbf{Procedural Consistency}: Applying this immutable codebook across both corpora ensures procedural consistency while preserving domain characteristics.

\subsection{Three-Level Analytical Protocol: Distinct Dimensions of Construct Validity}
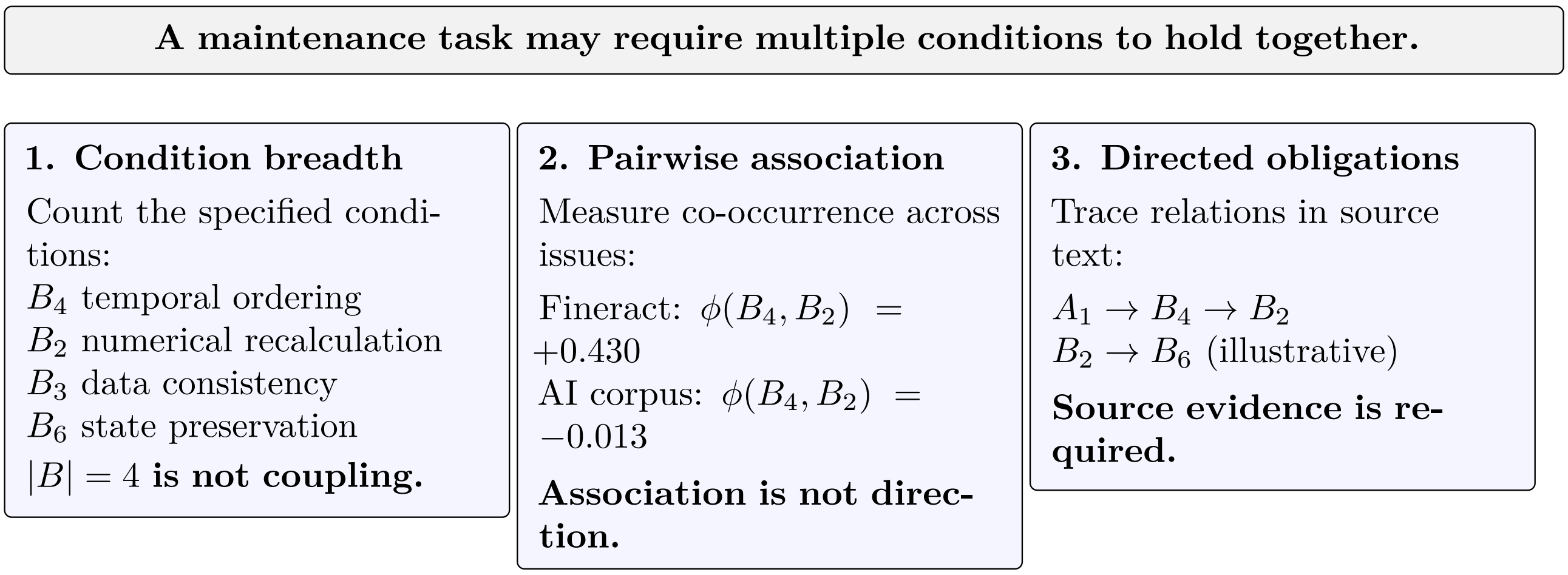
\begin{figure}[htbp]
\centering
\resizebox{\linewidth}{!}{%
\begin{tikzpicture}[
    box/.style={draw, rounded corners=2pt, fill=blue!4,
                text width=4.55cm, align=left, font=\small,
                inner sep=6pt, anchor=north west},
    root/.style={draw, rounded corners=2pt, fill=gray!10,
                text width=15.0cm, align=center, font=\small,
                inner sep=5pt, anchor=north west}
]
\node[root] at (0,0) {\textbf{A maintenance task may require multiple conditions to hold together.}};
\node[box] at (0,-1.15) {
\textbf{1. Condition breadth}\\[3pt]
Count the specified conditions:\\
$B_4$ temporal ordering\\
$B_2$ numerical recalculation\\
$B_3$ data consistency\\
$B_6$ state preservation\\[2pt]
\textbf{$|B|=4$ is not coupling.}
};
\node[box] at (5.05,-1.15) {
\textbf{2. Pairwise association}\\[3pt]
Measure co-occurrence across issues:\\[3pt]
Fineract: $\phi(B_4,B_2)=+0.430$\\
AI corpus: $\phi(B_4,B_2)=-0.013$\\[4pt]
\textbf{Association is not direction.}
};
\node[box] at (10.10,-1.15) {
\textbf{3. Directed obligations}\\[3pt]
Trace relations in source text:\\[3pt]
$A_1\rightarrow B_4\rightarrow B_2$\\
$B_2\rightarrow B_6$ (illustrative)\\[4pt]
\textbf{Source evidence is required.}
};
\end{tikzpicture}%
}
\caption{Three complementary lenses for multi-condition fulfillment: breadth, co-occurrence, and source-grounded directionality. The illustrated dependency chain synthesizes FINERACT-753 and FINERACT-2338 rather than quoting a single task; oracle inspection is a downstream application.}
\label{fig:analytical_framework}
\end{figure}

To separate the breadth of completion conditions from their relationships, we use three complementary lenses (Figure~\ref{fig:analytical_framework}): (1)~\textbf{Condition Breadth ($|B|$)} counts distinct tagged fulfillment criteria in a task, without asserting that they interact; (2)~\textbf{Corpus Co-occurrence ($\phi$)} measures pairwise association across issue specifications, without identifying direction or causality; and (3)~\textbf{Source-Grounded Directional Dependencies} traces action--condition or condition--condition obligations when supported by requirement text. The figure combines findings from FINERACT-753 and FINERACT-2338 as an illustrative synthesis, not a verbatim single-issue specification. Subsequent inspection of selected benchmark assertions is an optional evaluation-facing application of these three lenses, not a fourth measurement of task complexity.

\section{E1: Unmasking Semantic Heterogeneity Beyond Surface Metrics}

\subsection{Unmasking Semantic Heterogeneity Beyond Surface Churn}
Conventional benchmark leaderboards and dataset summaries commonly categorize maintenance tasks under broad native issue classifications (e.g., "Bug Fix") alongside summary metrics of code churn and file counts. To empirically evaluate whether tasks sharing comparable surface characteristics exhibit comparable requirement structures, we constructed a controlled matched-pair analysis across the 944 benchmark tasks.

We identified pairs of tasks $(t_i, t_j)$ that satisfied three equivalence constraints:
\begin{enumerate}
  \item Identical native repository issue classification.
  \item Identical modified file count ($|Files(t_i)| = |Files(t_j)|$).
  \item Code churn: \( |LOC_i-LOC_j| \le \max\{3,0.30\max(LOC_i,LOC_j)\} \).
\end{enumerate}

For each matched pair, we computed the Jaccard distance over their active fulfillment condition sets ($B(t_i), B(t_j)$):
$$d_J(t_i, t_j) = 1 - \frac{|B(t_i) \cap B(t_j)|}{|B(t_i) \cup B(t_j)|}$$

We emphasize that this matched-pair construction controls for surface task similarity (repository issue classification, file modification count, and line churn magnitude), but does not imply or guarantee identical cognitive problem-solving difficulty. Rather, it isolates whether tasks with matched issue classifications and file counts and churn within tolerance share common fulfillment criteria. Across $N=1,177$ surface-matched pairs (714 SWE-bench Pro pairs and 463 PolyBench Java pairs, spanning 581 unique tasks), the mean Jaccard distance is \textbf{$\bar{d}_J = 0.8138$} (median $1.0000$, sample standard deviation $\sigma = 0.2745$). Crucially, \textbf{57.4\% of all controlled pairs} (676 pairs) exhibit $d_J = 1.0000$, meaning their underlying fulfillment conditions are \textbf{completely disjoint} despite matching issue classifications and modified-file counts, with code churn within the specified tolerance.

A representative example is observed in \texttt{internetarchive/openlibrary}: OpenLibrary \#0143 (29 LOC, 1 file, tagged Bug Fix) modifies author-edition entity relationships requiring relational state consistency ($A_1 \rightarrow B_3, B_5$), whereas OpenLibrary \#0233 (27 LOC, 1 file, tagged Bug Fix) corrects string sanitization and character encoding compatibility ($A_1 \rightarrow B_1, B_6, B_{11}$). Although both tasks have comparable surface characteristics (1 modified file each, 27 and 29 LOC, native 'Bug Fix' tag), their underlying semantic fulfillment conditions are completely disjoint ($d_J = 1.0000$), demonstrating that surface metadata fails to reflect whether a task demands relational database integrity (\#0143) or character encoding transformations (\#0233).

\subsection{Stratified Benchmark Family Profile (N=944)}
Rather than aggregating AI benchmarks into a monolithic entity, we analyze empirical distributions stratified across the three constituent benchmark families in Table~\ref{tab:stratified_distribution} (\path{gate3_stratified_family_metrics.csv}).

\begin{table}[htbp]
\centering
\footnotesize
\renewcommand{\arraystretch}{0.88}
\setlength{\tabcolsep}{3.5pt}
\caption{Stratified 18-Criteria Distribution Across Benchmark Families ($N=944$).}
\label{tab:stratified_distribution}
\begin{tabularx}{\linewidth}{l X c c c c}
\toprule
\textbf{Code} & \textbf{Codebook Criterion Name} & \textbf{AI All ($N=944$)} & \textbf{Pro ($N=731$)} & \textbf{Poly ($N=165$)} & \textbf{EVO ($N=48$)} \\
\midrule
\textbf{A1} & Product Operation \& Business Rules & 426 (45.13\%) & 300 (41.04\%) & 84 (50.91\%) & 42 (87.50\%) \\
\textbf{A2} & Structure \& Implementation Method & 572 (60.59\%) & 505 (69.08\%) & 50 (30.30\%) & 17 (35.42\%) \\
\textbf{A3} & Testing \& Quality Assurance & 94 (9.96\%) & 70 (9.58\%) & 10 (6.06\%) & 14 (29.17\%) \\
\textbf{A4} & Execution \& Operations & 122 (12.92\%) & 93 (12.72\%) & 20 (12.12\%) & 9 (18.75\%) \\
\textbf{A5} & Deployment \& Environment Migration & 60 (6.36\%) & 39 (5.34\%) & 9 (5.45\%) & 12 (25.00\%) \\
\textbf{A6} & Development Process \& Docs & 53 (5.61\%) & 25 (3.42\%) & 17 (10.30\%) & 11 (22.92\%) \\
\midrule
\textbf{B1} & Functional Result \& Input Condition & 315 (33.37\%) & 276 (37.76\%) & 25 (15.15\%) & 14 (29.17\%) \\
\textbf{B2} & Numerical \& Calculation Precision & 65 (6.89\%) & 49 (6.70\%) & 11 (6.67\%) & 5 (10.42\%) \\
\textbf{B3} & Data Relationship \& Consistency & 214 (22.67\%) & 184 (25.17\%) & 25 (15.15\%) & 5 (10.42\%) \\
\textbf{B4} & Temporal Semantics \& Ordering & 86 (9.11\%) & 77 (10.53\%) & 7 (4.24\%) & 2 (4.17\%) \\
\textbf{B5} & State Transition \& Business Workflow & 131 (13.88\%) & 109 (14.91\%) & 20 (12.12\%) & 2 (4.17\%) \\
\textbf{B6} & Preservation of Existing Data \& Behavior & 368 (38.98\%) & 336 (45.96\%) & 17 (10.30\%) & 15 (31.25\%) \\
\textbf{B7} & Concurrency \& Transaction Boundaries & 42 (4.45\%) & 35 (4.79\%) & 4 (2.42\%) & 3 (6.25\%) \\
\textbf{B8} & Authorization \& Data Protection & 154 (16.31\%) & 130 (17.78\%) & 16 (9.70\%) & 8 (16.67\%) \\
\textbf{B9} & Performance \& Resource Utilization & 83 (8.79\%) & 60 (8.21\%) & 13 (7.88\%) & 10 (20.83\%) \\
\textbf{B10} & Service Continuity \& Failure Handling & 87 (9.22\%) & 72 (9.85\%) & 8 (4.85\%) & 7 (14.58\%) \\
\textbf{B11} & Interface \& Environmental Compatibility & 391 (41.42\%) & 311 (42.54\%) & 51 (30.91\%) & 29 (60.42\%) \\
\textbf{B12} & User Interaction Semantics & 183 (19.39\%) & 169 (23.12\%) & 4 (2.42\%) & 10 (20.83\%) \\
\bottomrule
\end{tabularx}
\end{table}

\subsection{Multi-Attribute Co-presence Complexity (\texorpdfstring{$|B|$}{|B|})}
Table~\ref{tab:coupling_complexity} reports the distribution of $|B|$ (\path{gate3_coupling_complexity.csv}).

\begin{table}[htbp]
\centering
\footnotesize
\renewcommand{\arraystretch}{0.88}
\setlength{\tabcolsep}{3.5pt}
\caption{Fulfillment-Condition Breadth: Distribution of $|B|$ (not a measure of dependency strength).}
\label{tab:coupling_complexity}
\begin{tabularx}{\linewidth}{X c c c c c}
\toprule
\textbf{Metric / Bin} & \textbf{Fineract ($N=2,925$)} & \textbf{AI All ($N=944$)} & \textbf{Pro ($N=731$)} & \textbf{Poly ($N=165$)} & \textbf{EVO ($N=48$)} \\
\midrule
\textbf{Mean $|B|$} & \textbf{1.78} (1.7798) & \textbf{2.24} (2.2447) & \textbf{2.47} (2.4733) & \textbf{1.22} (1.2182) & \textbf{2.29} (2.2917) \\
\textbf{Median $|B|$} & 2.0 & 2.0 & 2.0 & 1.0 & 2.0 \\
\midrule
$|B| = 0$ & 331 (11.32\%) & 3 (0.32\%) & 0 (0.00\%) & 3 (1.82\%) & 0 (0.00\%) \\
$|B| = 1$ & 872 (29.81\%) & 248 (26.27\%) & 109 (14.91\%) & 126 (76.36\%) & 13 (27.08\%) \\
$|B| = 2$ & 1,097 (37.50\%) & 357 (37.82\%) & 304 (41.59\%) & 33 (20.00\%) & 20 (41.67\%) \\
$|B| = 3$ & 451 (15.42\%) & 231 (24.47\%) & 219 (29.96\%) & 3 (1.82\%) & 9 (18.75\%) \\
$|B| = 4$ & 112 (3.83\%) & 79 (8.37\%) & 77 (10.53\%) & 0 (0.00\%) & 2 (4.17\%) \\
$|B| = 5$ & 37 (1.26\%) & 15 (1.59\%) & 12 (1.64\%) & 0 (0.00\%) & 3 (6.25\%) \\
$|B| = 6$ & 21 (0.72\%) & 6 (0.64\%) & 6 (0.82\%) & 0 (0.00\%) & 0 (0.00\%) \\
$|B| = 7$ & 4 (0.14\%) & 3 (0.32\%) & 2 (0.27\%) & 0 (0.00\%) & 1 (2.08\%) \\
$|B| = 8$ & 0 (0.00\%) & 2 (0.21\%) & 2 (0.27\%) & 0 (0.00\%) & 0 (0.00\%) \\
\midrule
\textbf{Subtotal ($|B| \ge 2$)} & \textbf{1,722 (58.87\%)} & \textbf{693 (73.41\%)} & \textbf{622 (85.09\%)} & \textbf{36 (21.82\%)} & \textbf{35 (72.92\%)} \\
\bottomrule
\end{tabularx}
\end{table}

\subsection{Attribute Breadth Is Not Joint Fulfillment Structure}
In the analyzed corpora, SWE-bench Pro has a higher mean number of tagged fulfillment conditions ($|B|=2.47$) than Apache Fineract ($1.78$, Welch's $t=15.78, p < 10^{-50}$). Thus the contrast cannot be reduced to a claim that benchmark tasks contain fewer conditions. But $|B|$ measures breadth only: two tasks with the same count may differ in which conditions must hold jointly and whether one condition triggers another. This result does not by itself show that enterprise issues are universally deeper or more difficult. 

This motivates the next question: how do specified conditions combine? Section 5 first examines corpus-level pairwise association and then traces source-grounded directional relationships in enterprise issues. Selected benchmark cases subsequently illustrate why the distinction can matter for evaluation, without treating test-assertion coverage as our principal empirical outcome.

\section{E2: Joint Fulfillment Patterns and Directional Dependencies}

\subsection{Macro Statistical Co-occurrence Divergence (\texorpdfstring{$\phi$}{phi})}
To measure how often pairs of fulfillment conditions are specified together, we compute Pearson's $\phi$ across the studied corpora (Table~\ref{tab:phi_correlation}; \texttt{gate3\_phi\_correlation\_matrix.csv}). This is a measure of association, not evidence that one condition triggers another. The near-zero $\phi$ for $B_4$ and $B_2$ in the pooled AI tasks ($-0.013$, $p=0.825$) indicates little observed pairwise association, not proof of independence or the absence of dependent tasks.

\noindent\textbf{Conditional and higher-order co-specification.} To make the joint-condition contrast interpretable, we calculate the proportion of $B_4$-tagged issues that also specify $B_2$. In Fineract, this is 254/520 (48.85\%); in the pooled AI benchmark corpus, it is 5/86 (5.81\%). The descriptive conditional risk ratio is 8.40 (95\% CI: 3.57--19.76; two-sided Fisher's exact test, $p=3.14\times10^{-16}$). An exploratory higher-order check finds $B_4$ and $B_2$ jointly specified with at least one of $B_3$ (data consistency) or $B_6$ (preservation) in 108/2,925 Fineract issues versus 1/944 benchmark tasks. These counts are computed from the frozen criterion annotations (\texttt{gate5\_multi\_condition\_summary.csv}); they measure co-specification, \emph{not} directed dependency rates or causal depth. Because the corpora represent different software domains, the contrast cannot be attributed to benchmark design alone.

To examine directionality, Section~\ref{subsec:micro_causal} separately inspects source-grounded requirements.

\begin{table}[htbp]
\centering
\footnotesize
\renewcommand{\arraystretch}{0.88}
\setlength{\tabcolsep}{3.5pt}
\caption{Pairwise Co-occurrence (\texorpdfstring{$\phi$}{phi}); associations do not establish causal direction.}
\label{tab:phi_correlation}
\begin{tabularx}{\linewidth}{l X r r}
\toprule
\textbf{Pair} & \textbf{Condition Pair (non-directional)} & \textbf{Fineract ($N=2,925$)} & \textbf{AI Bench ($N=944$)} \\
 & & \textbf{Co-occur / $\phi$} & \textbf{Co-occur / $\phi$ ($p$)} \\
\midrule
$B_4 \leftrightarrow B_2$ & Temporal / Numerical & 254 / $\mathbf{+0.430}$ & 5 / $\mathbf{-0.013}$ ($0.825$) \\
$A_1 \leftrightarrow B_4$ & Business Rule / Temporal & 471 / $\mathbf{+0.309}$ & 42 / $\mathbf{+0.024}$ ($0.496$) \\
$A_1 \leftrightarrow B_2$ & Business Rule / Numerical & 440 / $\mathbf{+0.343}$ & 34 / $\mathbf{+0.039}$ ($0.246$) \\
$A_1 \leftrightarrow B_3$ & Business Rule / Consistency & 538 / $\mathbf{+0.329}$ & 115 / $\mathbf{+0.094}$ ($0.005$) \\
$B_4 \leftrightarrow B_3$ & Temporal / Consistency & 138 / $\mathbf{+0.069}$ & 22 / $\mathbf{+0.022}$ ($0.501$) \\
$B_4 \leftrightarrow B_6$ & Temporal / Preservation & 84 / $\mathbf{+0.068}$ & 28 / $\mathbf{-0.042}$ ($0.246$) \\
$B_2 \leftrightarrow B_3$ & Numerical / Integrity & 189 / $\mathbf{+0.226}$ & 14 / $\mathbf{-0.007}$ ($1.000$) \\
$B_2 \leftrightarrow B_6$ & Numerical / Preservation & 64 / $\mathbf{+0.036}$ & 19 / $\mathbf{-0.054}$ ($0.113$) \\
$B_3 \leftrightarrow B_6$ & Integrity / Preservation & 89 / $\mathbf{+0.053}$ & 63 / $\mathbf{-0.106}$ ($0.001$) \\
$B_7 \leftrightarrow B_6$ & Isolation / Preservation & 8 / $\mathbf{+0.005}$ & 14 / $\mathbf{-0.025}$ ($0.519$) \\
$B_7 \leftrightarrow B_3$ & Isolation / Integrity & 20 / $\mathbf{+0.040}$ & 7 / $\mathbf{-0.031}$ ($0.451$) \\
\bottomrule
\end{tabularx}
\end{table}

\subsection{Source-Grounded Directional Fulfillment in Enterprise Maintenance}\label{subsec:micro_causal}
To characterize what a joint completion obligation can look like, we examine \textbf{16 source-grounded business requirement relationships} in Apache Fineract (Table~\ref{tab:fineract_dependencies}; \texttt{gate2\_dependencies.csv}). The FINERACT-2338 examples connect backdating to repayment recalculation and preservation constraints; FINERACT-2836 specifies transaction boundaries for savings-charge processing. Supporting quotations were checked against source text (\texttt{gate2\_raw\_source\_snippets.json}). These purposively documented relationships illustrate directionality; they are not a frequency estimate or evidence that all Fineract issues exhibit this structure.

\begin{table}[htbp]
\centering
\small
\caption{Representative Source-Grounded Requirement Dependencies in Apache Fineract Maintenance (Full set of 16 documented relationships in the replication package).}
\label{tab:fineract_dependencies}
\begin{tabularx}{\linewidth}{c c l c X}
\toprule
\textbf{ID} & \textbf{Issue} & \textbf{Classification} & \textbf{Structural Chain} & \textbf{Verbatim Jira Excerpt} \\
\midrule
\multicolumn{5}{l}{\textbf{Causal Cascade Focus: Multi-Tier Loan Modification (FINERACT-2338)}} \\
\textbf{DEP-01} & FINERACT-2338 & Explicit Dependency & $A_1 \rightarrow B_4$ & ``Backdate can be done on already paid Installments as well'' \\
\textbf{DEP-02} & FINERACT-2338 & Explicit Dependency & $B_4 \rightarrow B_2$ & ``Repayment schedule will be recalculated with New EMI and Interest from the Interest applied schedule date'' \\
\textbf{DEP-03} & FINERACT-2338 & Explicit Dependency & $B_2 \rightarrow B_6$ & ``No of Installments will remain the same, only EMI and Interest would get affected'' \\
\textbf{DEP-04} & FINERACT-2338 & Explicit Dependency & $B_4 \rightarrow B_3$ & ``standard accounting rules and not charge-off accounting rules'' \\
\midrule
\multicolumn{5}{l}{\textbf{Cross-Entity Invariant Preservation \& Transaction Isolation}} \\
\textbf{DEP-05} & FINERACT-753 & Explicit Dependency & $A_1 \rightarrow B_4, B_2$ & ``now reverse the deposit of 500 .'' \\
\textbf{DEP-07} & FINERACT-2836 & Explicit Dependency & $B_7 \rightarrow B_6$ & ``Each due savings-charge installment is processed in an independent transaction.'' \\
\bottomrule
\end{tabularx}
\end{table}

\subsection{Illustrative Evaluation Implications in Four Benchmark Tasks}\label{subsec:oracle_eval}
The preceding results concern \emph{specified conditions and their relationships}, not agent success or general test quality. As a secondary illustration, we inspected task text, source diffs, and selected test assertions for four benchmark tasks (Table~\ref{tab:comparative_cases}; \texttt{data\_and\_tables/ai\_case\_sources/}). These cases show why a multi-condition specification can call for more than a localized check, and also show instances of appropriate alignment. They are not representative samples and do not establish full-harness behavior.

We establish a unified \textbf{3-Question Comparative Protocol} applied systematically across all four axes: (1)~\textbf{Specification Mandate ($R_{\text{spec}}$)}: operational state transformations and invariants required by the raw task description and DoD; (2)~\textbf{Oracle Assertion Scope ($R_{\text{eval}}$)}: specific conditions directly asserted by test code in the benchmark's evaluation patch; and (3)~\textbf{Construct Alignment or Structural Gap}: structural correspondence or divergence between mandate and assertions. To distinguish task demands from implementation details, requirement conditions are categorized into three provenance tiers: (a)~\textbf{Explicit Requirement} (directly stated in prompt or interface); (b)~\textbf{Developer-Implemented Obligation} (concrete implementation in reference patch); and (c)~\textbf{Inferred Requirement} (deduced from architectural context but unstated in prompt).

\begin{table}[htbp]
\centering
\scriptsize
\caption{Comparative Case Analysis: Test Oracle Construct Coverage Across Four Qualitative Benchmark Axes (Unabridged matrix in replication package).}
\label{tab:comparative_cases}
\begin{tabularx}{\linewidth}{@{} p{1.5cm} X X p{1.3cm} X @{}}
\toprule
\textbf{Axis / Task} & \textbf{Specification Mandate ($R_{\text{spec}}$)} & \textbf{Evaluation Test Oracle ($R_{\text{eval}}$)} & \textbf{Observed Scope} & \textbf{Construct Alignment or Gap Analysis} \\
\midrule
\textbf{Axis 1 (Direct)}\newline NodeBB \#0030 & 
\textbf{Explicit ($A_5 \rightarrow B_6$)}: DB migration script (\texttt{chat\_allow\_list.js}, 44 LOC) migrating legacy \texttt{restrictChat} records. & 
\textbf{Leaf Unit Assertion}: Asserts \texttt{canMessageUser()} on mock objects in memory (\path{test/messaging.js}, L103--121). & 
In-memory leaf & 
\textbf{Construct Underrepresentation}: Direct test patch asserts leaf permissions; DB migration script unasserted in analyzed test patch. \\
\midrule
\textbf{Axis 2 (Scope)}\newline Element-Web \#0139 & 
\textbf{Explicit badge update; inferred cache persistence}: Recalculate unread badges; IndexedDB cache synchronization is inferred rather than explicitly mandated. & 
\textbf{Mocked State}: Asserts helper return values under synthetic event mocks (\texttt{test/Unread-test.ts}). & 
In-memory mock & 
\textbf{Mocked Logic vs. Persistent Cache}: In-memory transitions verified; asynchronous IndexedDB cache persistence outside evaluation scope. \\
\midrule
\textbf{Axis 3 (Hierarchy)}\newline RocketMQ \#0819 & 
\textbf{Explicit ($B_7, B_2$)}: Thread-safe counter insertion (\texttt{StatsItemSet}, 20kL) under JVM thread contention. & 
\textbf{Thread Stress Test}: 100 worker threads executing 10,000 tasks (\texttt{StatsItemSetTest.java}). & 
JVM thread contention & 
\textbf{Single JVM Concurrency Met}: Checks concurrent counter accumulation within a single JVM; distributed transaction boundaries are not mandated by the task. \\
\midrule
\textbf{Axis 4 (Aligned)}\newline Navidrome \#0495 & 
\textbf{Explicit ($A_4, B_6, B_{10}$)}: Online SQLite DB backup, destructive DDL recovery, and retention pruning. & 
\textbf{Integrated Engine Test}: Executes backup, runs destructive DDL, restores DB, asserts pruning (\texttt{db/backup\_test.go}). & 
Persistent DB state & 
\textbf{Aligned Invariant Testing}: Positive case in which the inspected tests directly address the explicitly required persistent-state operations. \\
\bottomrule
\end{tabularx}
\end{table}

\subsubsection{Comparative Case Analysis across the Three Questions and Provenance Tiers}
Applying the protocol illustrates different alignments between explicit task obligations and the inspected assertions; it does not quantify the prevalence of evaluation gaps:

\noindent\textbf{Axis 1 (NodeBB \#0030: Construct Underrepresentation).} \textit{Mandate ($R_{\text{spec}}$)}: The issue explicitly mandates an upgrade path in \texttt{src/upgrades/4.3.0/chat\_allow\_list.js} to migrate legacy \texttt{restrictChat=1} records (\textbf{Explicit Requirement}), implemented via a 44-line batch migration script (\texttt{batch.processSortedSet}, 500 records/chunk) in the developer patch (+383/-50 lines across 12 files). \textit{Oracle ($R_{\text{eval}}$)}: Static inspection of the analyzed test patch (\texttt{test/messaging.js:L103-121}) confirms it creates mock users in memory with pre-populated \texttt{chatAllowList} arrays, asserting \texttt{Messaging.canMessageUser()} with zero assertions exercising \texttt{src/upgrades/}. \textit{Structural Gap}: Static comparison confirms developer migration omission in an ablated patch modifying 11 files, coupled with complete migration coverage absence in analyzed test assertions. Within this verified static inspection scope, assertions evaluate localized leaf permissions while leaving the database migration unasserted.

\noindent\textbf{Axis 2 (Element-Web \#0139: Mocked Logic vs. Inferred Cache Persistence).} \textit{Mandate ($R_{\text{spec}}$)}: The issue mandates recomputing unread badges when receipts arrive across thread timelines, implemented in \texttt{src/Unread.ts} (\textbf{Explicit Requirement} and \textbf{Developer Obligation}). Persistent IndexedDB cache synchronization across network streams is an \textbf{Inferred Requirement}. \textit{Oracle ($R_{\text{eval}}$)}: Test code in \texttt{test/Unread-test.ts} asserts in-memory helper returns under synthetic mocked events (\texttt{mocked(haveRendererForEvent)}). \textit{Structural Gap}: While in-memory state transitions for the Explicit Requirement are validated, persistent IndexedDB synchronization and asynchronous event reordering remain unasserted, reflecting the boundary between unit mock verification and inferred persistent client state.

\noindent\textbf{Axis 3 (RocketMQ \#0819: JVM Contention vs. Transaction Boundaries).} \textit{Mandate ($R_{\text{spec}}$)}: The task mandates thread-safe counter accumulation under concurrent thread contention in \texttt{StatsItemSet} (\textbf{Explicit Requirement}), resolved using \texttt{putIfAbsent} (\textbf{Developer Obligation}); distributed transactions were neither requested nor implied. \textit{Oracle ($R_{\text{eval}}$)}: Test code in \texttt{StatsItemSetTest.java} spawns 100 worker threads executing 10,000 tasks, asserting atomic accumulation of 20,000L. \textit{Structural Alignment}: The inspected tests exercise concurrent accumulation within a single JVM, in line with the Explicit Requirement.

\noindent\textbf{Axis 4 (Navidrome \#0495: Aligned Invariant Testing Counterexample).} \textit{Mandate ($R_{\text{spec}}$)}: The task explicitly mandates online SQLite database backup, destructive DDL recovery, and backup retention pruning (\textbf{Explicit Requirement}), implemented across \texttt{cmd/backup.go} and \texttt{db/backup.go} (\textbf{Developer Obligation}). \textit{Oracle ($R_{\text{eval}}$)}: Test code in \texttt{db/backup\_test.go} creates real SQLite backups, executes destructive DDL (\texttt{DELETE FROM sqlite\_master}), verifies schema restoration, and asserts retention pruning policies. \textit{Structural Alignment}: The inspected tests provide direct evidence of alignment with the explicitly specified SQLite backup, schema restoration, and retention requirements.

\section{Threats to Validity}\label{sec:threats}

\subsection{Internal \& Construct Validity}
A primary threat to internal and construct validity is potential subjective bias in applying the 18-criteria taxonomy. Semantic coding across 944 AI benchmark tasks was conducted by a single investigator against frozen codebook definitions rather than a multi-researcher adjudication committee. While all 3,446 criteria assignments were validated through automated scripts (\texttt{validate\_*.py}), programmatic offset checking guarantees textual quotation fidelity and source hash integrity rather than multi-coder consensus; we make no claim of dual-coder agreement statistics. We mitigated potential subjective drift through four procedural controls: (1)~freezing the 18-criteria operational definitions and boundary rules prior to corpus-wide annotation (\texttt{framework-snapshot.json}), anchoring criteria in domain-neutral standards (ISO/IEC/IEEE 14764, ISO/IEC 25010); (2)~requiring an explicit verbatim textual quote from the task specification for every criterion assignment; (3)~providing all evidence quotes, code diff references, and validation scripts in our replication package for independent audit; and (4)~verifying all 16 Fineract dependency chains via automated byte-level offset matching against raw Jira JSON archives (\texttt{gate2\_raw\_source\_snippets.json}). Additionally, to address potential matched-pair clustering in E1 across identical repositories, we confirmed that semantic heterogeneity persists across cross-repository pairs regardless of repository origin.

\subsection{External Validity \& Scope Boundaries}
Evaluating 944 benchmark tasks across three established benchmarks spanning multiple languages (Python, Java, TypeScript, Go), we explicitly delimit our empirical scope: (1)~\textbf{Empirical Scope vs. Universal Generalization}: Findings reflect quantitative replicability and static verification across the analyzed corpora, rather than a universal claim that all AI benchmarks lack causal dependencies; well-designed harnesses can evaluate persistent state operations (Navidrome \#0495); (2)~\textbf{Static Oracle Analysis vs. Dynamic Solve Rates}: We examine test oracle construct representation via diffs and assertions, without claiming dynamic execution pass rates for unexecuted models; (3)~\textbf{Domain Confounding}: Fineract is a banking system with temporally dependent ledger requirements, while the benchmark corpus spans different projects. Using a shared ISO-informed taxonomy makes labels comparable but does not control for domain composition or demonstrate benchmark selection bias. This also applies to the conditional and exploratory higher-order co-specification rates, whose sparse benchmark cells limit precision; and (4)~\textbf{Dependency and Case Selection}: The 16 source-grounded relationships and four oracle case studies illustrate possible structures, not their prevalence in the populations. We do not establish that enterprise maintenance universally requires more dependencies, measure production failure rates, or assess a complete execution harness from an inspected test patch.

\section{Discussion \& Benchmark Design Implications}

\subsection{Re-evaluating the Benchmark Complexity Paradigm}
Our findings demonstrate that software maintenance complexity cannot be adequately characterized solely by surface metadata such as lines of code churn or file navigation graphs. In E1, 1,177 pairs matched on native issue classification and modified-file count, with code churn within the prespecified tolerance, exhibited mean fulfillment-set Jaccard distance 0.8138; 57.4\% (676 pairs) had completely disjoint condition sets ($d_J=1.0000$). This indicates that matching on native issue classification, exact modified-file counts, and code churn within the specified tolerance does not ensure semantic similarity in fulfillment conditions.

Furthermore, SWE-bench Pro has higher mean attribute breadth ($|B|=2.47$) than Apache Fineract ($1.78$) in the studied corpora. Thus \textbf{the number of tagged fulfillment conditions alone cannot characterize maintenance requirement structure}; their co-specification patterns and source-grounded dependencies must also be considered. The Fineract examples show how a backdated operation can impose linked recalculation and preservation obligations. The co-occurrence contrast and the additional conditional rates ($P(B_2\mid B_4)$: 48.85\% in Fineract versus 5.81\% in the pooled AI corpus) are consistent with different patterns of jointly specified conditions. They do not measure causal dependency strength, and domain composition prevents attributing this contrast to benchmark design alone. Selected oracle inspections illustrate implications for checking explicit obligations rather than forming the principal empirical claim.

\subsection{Implications for Describing and Evaluating Maintenance Tasks}
The empirical results motivate three reporting and design considerations, rather than a claim that any one test architecture is universally superior:

\noindent\textbf{1.~Report fulfillment-condition profiles alongside surface size}: Task descriptions and benchmark reports should distinguish native issue labels, file counts, and churn from the types and breadth ($|B|$) of required completion conditions. In E1, superficially matched tasks frequently have different condition sets.

\noindent\textbf{2.~Represent linked conditions when the specification warrants them}: Beyond listing attributes, descriptions should preserve source-grounded dependencies, such as a backdated change that triggers financial recalculation while retaining historical consistency. The observed $\phi$ differences motivate closer analysis but do not themselves prove directional dependencies or broader causal mechanisms.

\noindent\textbf{3.~Align checks with explicitly required multi-condition completion}: If a task specifies a migration, transaction boundary, or state-preservation obligation, evaluation design should consider checking that obligation jointly with the local functional change. NodeBB and Navidrome illustrate contrasting scopes in inspected patches. This is a downstream implication of requirement analysis, not evidence that benchmark oracles generally fail. When possible, benchmark reports should make these verification boundaries transparent.

\section{Conclusion}

Using 944 analyzed coding-benchmark tasks and 2,925 Apache Fineract issues, we examined how maintenance completion conditions differ in breadth, co-occurrence, and source-grounded directionality.

\noindent\textbf{RQ1 (Semantic Heterogeneity and Breadth):} Among 1,177 pairs matched on issue type, modified-file count, and churn tolerance, mean fulfillment-set Jaccard distance was $0.8138$, with 57.4\% disjoint sets. SWE-bench Pro also had greater mean attribute breadth ($|B|=2.47$) than Fineract ($1.78$): an explanation based on fewer benchmark attributes is therefore insufficient for these corpora.

\noindent\textbf{RQ2a (Co-occurrence):} Temporal and numerical conditions exhibited positive association in Fineract ($B_4$/$B_2$, $\phi=+0.430$, 254 issues) but near-zero association in the pooled benchmark corpus ($\phi=-0.013$, $p=0.825$, 5 tasks). Among items specifying $B_4$, $B_2$ also appears in 254/520 Fineract issues (48.85\%) versus 5/86 benchmark tasks (5.81\%). These domain-confounded co-specification contrasts do not establish directional dependency rates or prove missing dependencies across AI benchmarks.

\noindent\textbf{RQ2b (Directional Fulfillment):} Sixteen source-grounded Fineract relationships illustrate how maintenance completion may depend on multiple linked obligations, including backdating, recalculation, and preservation constraints. Four selected benchmark cases show that the inspected evaluation assertions address explicit obligations with varying scope; these cases do not establish a general oracle-failure rate.

\textbf{Together, our results establish that matching surface task characteristics does not ensure similar fulfillment conditions and that condition counts alone do not characterize observed joint-specification patterns.} Maintenance benchmarks should therefore distinguish condition breadth, statistical association, and source-grounded directional obligations. Test-assertion alignment remains a downstream application rather than the principal empirical claim.

\section*{Data Availability}
Analysis datasets, selected source-grounded evidence, and targeted verification scripts are publicly available in the replication package: \href{https://github.com/HeeyongEun1982/ai-enterprise-maintenance-benchmark}{public GitHub repository}. Some upstream source materials and complete benchmark execution environments are not redistributed.

\bibliographystyle{unsrtnat}
\bibliography{references}
\end{document}